\documentclass[12pt]{iopart}

\usepackage[colorlinks,pdfusetitle,urlcolor=blue,citecolor=blue,linkcolor=blue,bookmarksnumbered,plainpages=false]{hyperref}

\usepackage{color}
\usepackage{soul}
\usepackage{amssymb}
\usepackage{amsfonts}

\usepackage{bm}
\usepackage{nicefrac}
\usepackage{siunitx}

\usepackage{graphicx}
\usepackage{bm}
\usepackage{braket}
\usepackage{mathrsfs}
\usepackage[normalem]{ulem}
\usepackage{iopams}
\usepackage{tikz}
\usetikzlibrary{positioning}

\expandafter\let\csname equation*\endcsname\relax
  \expandafter\let\csname endequation*\endcsname\relax
\usepackage{amsmath}
\usepackage{amssymb}
\usepackage{amsfonts}
\usepackage{yfonts}

\def\<{\langle}
\def\>{\rangle}
\newcommand{\tens}[1]{\mathbb{#1}}
\newcommand*\colvec[3][]{
		\begin{pmatrix}\ifx\relax#1\relax\else#1\\\fi#2\\#3\end{pmatrix}
	}
	\makeatletter
	\DeclareRobustCommand{\cev}[1]{%
	  \mathpalette\do@cev{#1}%
	}
	\newcommand{\do@cev}[2]{%
	  \fix@cev{#1}{+}%
	  \reflectbox{$\m@th#1\vec{\reflectbox{$\fix@cev{#1}{-}\m@th#1#2\fix@cev{#1}{+}$}}$}%
	  \fix@cev{#1}{-}%
	}
	\newcommand{\fix@cev}[2]{%
	  \ifx#1\displaystyle
		\mkern#23mu
	  \else
		\ifx#1\textstyle
		  \mkern#23mu
		\else
		  \ifx#1\scriptstyle
		    \mkern#22mu
		  \else
		    \mkern#22mu
		  \fi
		\fi
	  \fi
	}

	\makeatother
\definecolor{darkgreen}{rgb}{0.27451, 0.458824, 0.22745}
\usepackage{color}

\newcommand{\B}[1]{\mathbf{#1}}

\begin{document}

\setlength{\parindent}{0cm}

\title[Casimir effect for PEMCs: A sum rule for attractive/repulsive forces]{Casimir effect for perfect electromagnetic conductors (PEMCs): A sum rule for attractive/repulsive forces}

\author{Stefan Rode, Robert Bennett}

\address{Physikalisches Institut, Albert-Ludwigs-Universit\"at
Freiburg, Hermann-Herder-Str. 3, 79104 Freiburg, Germany}

\author{Stefan Yoshi Buhmann}

\address{Physikalisches Institut, Albert-Ludwigs-Universit\"at
Freiburg, Hermann-Herder-Str. 3, 79104 Freiburg, Germany}

\address{Freiburg Institute for Advanced Studies,
Albert-Ludwigs-Universit\"at Freiburg, Albertstr. 19, 79104 Freiburg,
Germany}

\begin{abstract}
We discuss the Casimir effect for boundary conditions involving perfect electromagnetic conductors (PEMCs). Based on the corresponding reciprocal Green's tensor we construct the Green's tensor for two perfectly reflecting plates with magnetoelectric coupling (non-reciprocal media) within the framework of macroscopic quantum electrodynamics. We calculate the Casimir force between two PEMC plates in terms of the PEMC parameter $M$ and the duality transformation angle $\theta$ resulting in a universal analytic expression that connects the attractive Casimir force with the repulsive Boyer force. We relate the results to the duality symmetry of electromagnetism. 
\end{abstract}

\maketitle

\section{Introduction}
Nonvanishing zero-point energies are a pervasive feature of quantum mechanics and quantum field theory.
The fact that energy fluctuations of the vacuum lead to physically observable macroscopic forces was first discovered by Hendrik Casimir in 1948 \cite{Casimir1948}, who calculated the attractive force between two uncharged metallic plates due to the fluctuations of the electromagnetic field, which turned out to be given by the simple expression
\begin{align}
f_\text{attr}=-\frac{\hbar c\pi^2}{240d^4} 
\end{align}
for plates separated by a distance $d$. The origin of this force is the non-vanishing expectation value of the squared electric field in the vacuum state, which is then modified by the presence of surfaces. These vacuum fluctuations give rise to various forms of matter-vacuum interaction. {The inverse fourth-power} distance-dependence leads to negligibly small forces on large distance scales. However, in the nanometre regime the Casimir effect and other vacuum fluctuation induced forces can become significant or even dominant. In particular,  the Casimir force poses a challenge for constructing microelectromechanical systems (MEMS) nanotechnology engineering \cite{Serry1995}. It causes effects such as stiction \cite{Tas1996, Buks2001}, which is the permanent adhesion of two nano-structural elements. In order to remove such impeding effects, possible ways of manipulating the Casimir force between bodies have been pursued.   \\
Of particular interest are repulsive Casimir forces \cite{Woods2016}. The first result in this field was obtained by Boyer in 1974 \cite{Boyer1974}, who considered an assembly of two parallel plates, one of them perfectly conducting, the other one perfectly permeable. He found the Casimir force to be repulsive in this case and showed that the ratio of his result to the attractive force calculated by Casimir reads 
\begin{align}
{f}_{\text{rep}}=-\frac 78{f}_{\text{attr}}.
\end{align} 
It has also been theoretically shown that the magnitude of the Casimir force between two plates of any magnetodielectric properties has to fall between the result of Casimir and the result of Boyer \cite{Henkel2017}.\\ Due to the difficulty of realising materials whose permeability is perfect or nearly perfect, other ways of implementing repulsive Casimir forces have been considered. Kenneth and Klich \cite{Kenneth2002} have discussed the opportunities of materials with non-trivial but finite magnetic susceptibilities for instance. As another approach the Casimir forces on materials with polarisation-twisting effects has been studied. In particular the vacuum interaction properties of topological insulators \cite{Grushin2011, Fuchs2017} and of chiral metamaterials \cite{Butcher2012, Rosa2008a},  have been investigated for generalised boundary conditions \cite{Asorey2013}. For the case of a scalar field confined by Robin boundary conditions, the Casimir force has been obtained to be either repulsive or attractive \cite{Romeo2002}. Here we will study perfect electromagnetic conductors (PEMCs) \cite{Lindell2005b,Sihvola2008a,Lindell2005,Lindell2005a}, an idealised class of nonreciprocal polarisation-mixing materials whose response is characterised by a single parameter $M$. We will calculate the Casimir force between two PEMC plates in terms of this parameter, which will allow us to quasi-tune the Casimir force between the two extremal values.  \\

From a more fundamental point of view, the Casimir force in PEMC media is of interest because of its close relation to duality invariance. It has been shown \cite{Buhmann2009} that a linear magnetodielectric medium breaks the duality invariance that holds for the free Maxwell equations, causing them to instead have a discrete $\mathbb{Z}_4$-symmetry. Allowing material response that violates Lorentz-reciprocity restores duality invariance \cite{Buhmann2012c} --- PEMC media provide these properties. For this reason we will determine the relation between the PEMC parameter $M$ and the duality angle of a perfect conductor to obtain a coherent picture of the impact of duality transformations on Casimir forces. 

\section{The Casimir force on nonreciprocal bodies}
The Green's tensor $\tens{G}(\B{r},\B{r}',\omega)=\tens{G}$ of Maxwell's equations in a region with tensor-valued permittivity $\varepsilon(\B{r},\omega)=\varepsilon$, permeability $\mu(\B{r},\omega)=\mu$ and cross-polarisabilities $\zeta(\B{r},\omega) = \zeta$ and $\xi(\B{r},\omega)=\xi$ (discussed in detail in Section \ref{BiIsoChap})  is defined to satisfy \cite{Buhmann2012c}; 
\begin{equation}\label{GTDef}
\left[\nabla \times \frac{1}{\mu} \star\nabla \times -\frac{\mathrm{i}\omega}{c}\nabla \times \frac{1}{\mu} \star \zeta  +\frac{\mathrm{i}\omega}{c}\xi \star \frac{1}{\mu} \star \nabla-\frac{\omega^2}{c^2} (\varepsilon -\xi \star \frac{1}{\mu} \star \zeta ) \right] \star\tens{G} =\tens{I} \delta(\B{r}-\B{r}')
\end{equation}
subject to appropriate boundary conditions, where $\star$ denotes spatial convolution 
\begin{equation}
[\tens{A}\star \tens{B}](\B{r},\B{r}')\equiv \int d^3\mathbf{s}\tens{A}(\mathbf{r},\mathbf{s})\cdot \tens{B}(\mathbf{s},\B{r}')\; . \end{equation}
and $\tens{I}$ is the identity matrix. Then, quantised electromagnetic fields can be constructed via \cite{Buhmann2012a} 
\begin{align}\label{Greens} {\hat{\mathbf{E}}}(\mathbf{r},\omega)=&\mathrm{i}\mu_0\omega\int d^3\mathbf{r}' \tens{G} (\mathbf{r},\mathbf{r}',\omega) \cdot{\hat{\mathbf{j}}_\text{N}}(\mathbf{r}',\omega)
\end{align}
where $\hat{\mathbf{j}}_\text{N}$ is a noise-current source, given explicitly by
\begin{align}
\hat{\mathbf{j}}_\text{N}(\mathbf{r},\omega)=&\sqrt{\frac{\hbar\omega}{\pi}}\tens{R}(\mathbf{r},\omega)\cdot\hat{\mathbf{f}}(\mathbf{r},\omega)
\end{align}
where $\tens{R}$ is a matrix satisfying $\tens{R} \cdot \tens{R}^\dagger=\frac{1}{2}(\tens{Q}+\tens{Q}^\dagger)$, with $\tens{Q}$ the conductivity tensor appearing in the generalised Ohm's law. The quantity $\hat{\mathbf{f}}$ is a bosonic excitation quasiparticle field satisfying;
\begin{gather}
[\mathbf{\hat{f}}(\mathbf{r},\omega),\mathbf{\hat{f}}^\dagger(\mathbf{r}',\omega')]=\delta(\mathbf{r}-\mathbf{r}')\delta(\omega-\omega'),
\label{eq42}
\end{gather}
with all other commutators being zero. From a macroscopic point of view the Casimir force $\mathbf{F}$ between arbitrary bodies can be interpreted as the ground-state expectation value of the Lorentz force, or equivalently by an integral over the Maxwell stress tensor $\hat{\tens{T}}$;
\begin{align}
\mathbf{F}=\int_{\partial V}\mathbf{dA}\cdot\langle {\hat{\tens{T}}}\rangle\label{eq8}
\end{align}
with 
\begin{align}
\label{eq39}\hat{\tens{T}}=\varepsilon_0\hat{\mathbf{E}}\otimes\hat{\mathbf{E}}+\frac{1}{\mu_0}\hat{\mathbf{B}}\otimes\hat{\mathbf{B}}-\frac 12 (\varepsilon_0\hat{\mathbf{E}}^2+\frac{1}{\mu_0}\hat{\mathbf{B}}^2)\mathbb{I}.
\end{align}
and the fields being obtained from Eq.~\eqref{Greens} together with $\hat{\B{B}} = (\mathrm{i}\omega)^{-1} \nabla \times \hat{\B{E}}$. 
We can now evaluate the expectation value in the vacuum state $\ket{\{0\}}$ of the noise current quanta $\hat{\mathbf{f}}$ by using $\hat{\B{f}}\ket{\{0\}}=0$ and the commutator \eqref{eq42} above. We will also use an integral relation that can be derived from the definition \eqref{GTDef} of the Green's tensor \cite{Buhmann2012c}
\begin{align}
\label{eq101}\textgoth{Im}[\tens{G}(\mathbf{r},\mathbf{r}',\omega)]=\mu_0\omega[\tens{G}(\omega)\star\tens{R}^\dagger(\omega)\tens{R}(\omega)\star\tens{G}^\dagger(\omega)](\mathbf{r},\mathbf{r}').
\end{align}
where 
\begin{equation}
\textgoth{Im}[\tens{A}]\equiv\frac{1}{2\mathrm{i}}(\tens{A}-\tens{A}^\dagger).
\end{equation}
is the generalised imaginary part of a tensor. Employing Eq.~\eqref{eq101} as well as the electric field given by Eq.~(\ref{Greens}), we can obtain the vacuum expectation values of the dyadic products appearing in Eq.~\eqref{eq39} in Fourier space:
\begin{align}
\nonumber \langle \hat{\mathbf{E}}(r,\omega)\otimes {\hat{\mathbf{E}}}(r',\omega')\rangle\!&=\!\frac{\mu_0^2\omega^3\hbar}{\pi}\!\!\int \!d^3\mathbf{s} \!\int\! d^3\mathbf{s}'\delta(\omega-\omega')\tens{G}(\mathbf{r},\mathbf{s},\omega)\!\cdot \!\tens{R}^\dagger(\mathbf{s},\omega)\tens{R}(\mathbf{s}',\omega)\!\cdot \!\tens{G}^\dagger(\mathbf{r}',\mathbf{s}',\omega)	\\
&=\frac{\mu_0\omega^2\hbar}{\pi}\textgoth{Im}[\tens{G}(\mathbf{r},\mathbf{r}',\omega)]\delta(\omega-\omega')\label{eq44},
\end{align}
and similar for all contributing terms in Eq.~\eqref{eq101}. Transforming back to position space and rotating to imaginary frequencies yields;
\begin{align}
\langle \mathbf{\hat{T}}\rangle =-\nonumber&\frac{\hbar}{2\pi}\int_0^\infty d\xi\int_{\partial V} dA\cdot\left\{\frac{\xi^2}{c^2}[\tens{G}^{(1)}(\B{r},\B{r},\mathrm{i}\xi)+\tens{G}^{(1)\text{T}}(\B{r},\B{r},\mathrm{i}\xi)]\right.\\
 &+\nonumber\left.\vec{\nabla}\times[\tens{G}^{(1)}(\B{r},\B{r}',\mathrm{i}\xi)+\tens{G}^{(1)\text{T}}(\B{r}',\B{r},\mathrm{i}\xi)]\times\cev{\nabla}'\right|_{\B{r}'\rightarrow \B{r}}\\
\nonumber&\quad-\frac 12\text{tr}\left[\frac{\xi^2}{c^2}[\tens{G}^{(1)}(\B{r},\B{r},\mathrm{i}\xi)+\tens{G}^{(1)\text{T}}(\B{r},\B{r},\mathrm{i}\xi)]\right.\\
&\left.\left.\left.\quad \quad+ \vec{\nabla}\times[\tens{G}^{(1)}(\B{r},\B{r}',\mathrm{i}\xi)+\tens{G}^{(1)\text{T}}(\B{r}',\B{r},\mathrm{i}\xi)]\times\cev{\nabla}'\right|_{\B{r}'\rightarrow \B{r}}\vphantom{\frac{\xi^2}{c^2}}\right]\mathbb{I}\vphantom{\frac{\xi^2}{c^2}}\right\}\label{eq54},
\end{align}
from which the force can be computed by means of Eq.~\eqref{eq8}. In this formula the Green's tensor has been replaced by its  scattering part $\tens{G}^{(1)}$ defined via
\begin{equation}
\tens{G}=\tens{G}^{(0)}+\tens{G}^{(1)}
\end{equation}
where $\tens{G}^{(1)}$ is the bulk part of the Green's tensor, which does not contribute to the Casimir force regardless of the system's geometry. In addition we exploit the fact that
\begin{align}
\lim_{|\omega|\rightarrow\infty}\tens{G}^{(0)}(\mathbf{r},\mathbf{r}')=\lim_{|\omega|\rightarrow\infty}\tens{G}(\mathbf{r},\mathbf{r}')=-\mathbb{I}\delta(\mathbf{r}-\mathbf{r}')
\end{align}
which functions as as a cutoff for high frequencies, allowing one to obtain a finite result. Note in particular that we did not assume the validity of the Lorentz reciprocity condition $\tens{G}(\mathbf{r},\mathbf{r}',\omega)=\tens{G}^\text{T}(\mathbf{r}',\mathbf{r},\omega)$, which is connected with time reversal invariance \cite{Buhmann2012c}. We hence have derived an expression for the Casimir force of arbitrary nonreciprocal bodies. 
\section{Bi-isotropic media and PEMCs}\label{BiIsoChap}
In order to obtain a tuneable Casimir force we will consider a class of materials whose reflection behaviour is in  some sense intermediate between the extreme cases of the perfect electric conductor (PEC) and perfect magnetic conductor (PMC), which are respectively characterised by infinite permittivity $\varepsilon$ or infinite permeability $\mu$. These materials are known as bi-isotropic, and in macroscopic quantum electrodynamics the response of such a medium is conveniently described by four material constants; the familiar $\varepsilon$ and $\mu$, as well as two cross-polarizabilities $\xi$ and $\zeta$. In principle all these quantities are permitted to be tensor-valued, which leads to the more general case of bi-anisotropic media. We will confine ourselves to bi-isotropic media, in which the material response shows no direction-dependence. This means that the four material constants are scalar-valued and fulfil the constitutive relations
\begin{align}
\label{eq01}\hat{\mathbf{D}}&=\varepsilon_0\varepsilon\hat{\mathbf{E}}+\frac{1}{c}\xi\hat{\mathbf{H}},\\
\label{eq02}\hat{\mathbf{B}}&=\mu_0\mu\hat{\mathbf{H}}+\frac 1c\zeta\hat{\mathbf{E}}.
\end{align} 
 For a fundamental theory of linear material response see Ref. \cite{Hehl2003}.
\subsection{Duality transformation}
By allowing for nonzero (or even infinite) cross-polarisabilities $\xi$ and $\zeta$ we achieve an interpolation between PECs and PMCs. To do this we note that Maxwell's equations for classical fields in media in the absence of free charges of currents can be arranged in the following way: 
\begin{align}
\nabla\cdot\colvec{Z_0{\mathbf{D}}}{\hat{\mathbf{B}}}&=0, \label{ConstRel1}\\
\nabla\times\colvec{{\mathbf{E}}}{Z_0{\mathbf{H}}}+\begin{pmatrix} 1 & 0 \\0 & -1 \end{pmatrix}\colvec{Z_0{\mathbf{D}}}{{\mathbf{B}}}&=0 \label{ConstRel2} ,
\end{align} 
where $Z_0=\sqrt{\mu_0/\varepsilon_0}$ denotes the impedance of free space. These equations are invariant under an SO(2) transformation, i.e. they remain valid when the vectors of fields are multiplied with a matrix of the form 
\begin{align}
\mathbb{D}=\begin{pmatrix} \cos(\theta)&\sin(\theta)\\-\sin(\theta) & \cos(\theta)\end{pmatrix}. 
\end{align}
The fields forming a vector in this formalism are called dual partners. The constitutive relations for the quantised fields then read \cite{Buhmann2009}
\begin{align}
\colvec{Z_0\hat{\mathbf{D}}}{\hat{\mathbf{B}}}=\frac 1c \begin{pmatrix} \varepsilon & \xi \\ \zeta & \mu\end{pmatrix} \colvec{\hat{\mathbf{E}}}{Z_0 \hat{\mathbf{H}}}+ \begin{pmatrix} 1 & \xi \\ 0 & \mu\end{pmatrix}\colvec{Z_0\mathbf{P}_\text{N}}{\mu_0\mathbf{M}_\text{N}},
\end{align}   
where the noise polarization $\hat{\mathbf{P}}_\text{N}$ and magnetisation polarization $\hat{\mathbf{M}}_\text{N}$ are related to the noise current $\hat{\B{j}}_\text{N}$. Note that in case of reciprocal materials the reduced number of degrees of freedom leads to the constraint that $\theta$ has to be a integer multiple of $\pi/2$, in which case the continuous symmetry of duality invariance hence reduces to a discrete $\mathbb{Z}_4$-symmetry \cite{Buhmann2009}. The consideration of polarisation of polarisation-mixing material constants $\xi$ and $\zeta$, however, restores the continuity of duality invariance \cite{Buhmann2012c}.\\

\subsection{Perfect electromagnetic conductors (PEMC)}     
We will now focus on perfect electromagnetic conductors (PEMC) as a special case of bi-isotropic materials. 
The concept of perfect electromagnetic conductors has been introduced by Lindell and others \cite{Lindell2005b,Lindell2005,Lindell2005a}, finding  applications in waveguide and antenna engineering \cite{Sihvola2008a}. At a boundary with normal vector $\mathbf{n}$ the PEMC reflection properties are defined via
\begin{align}
\label{eq03} \mathbf{n}\cdot(Z_0{\mathbf{D}}-M{\mathbf{B}})=& 0,\\
\label{eq04} \mathbf{n}\times(Z_0{\mathbf{H}}+M{\mathbf{E}})=& 0.
\end{align}
They show a transmission-free, polarisation-mixing reflection behaviour \cite{Lindell2005b}. The pseudoscalar material parameter $M$ is interpolates between PEC- ($M\rightarrow\infty$) and PMC ($M=0$ boundaries). 
We can now relate $M$ to the magnetoelectric material constants introduced in the previous section by comparing equations (\ref{eq03}) and (\ref{eq04}) with the general constitutive relations \eqref{ConstRel1} and \eqref{ConstRel2}. One arrives at
\begin{align}
\xi=\zeta=&\pm\sqrt{\mu\varepsilon}\label{eq62},\\
M=\frac{\xi}{\mu}=&\pm \sqrt{\frac{\varepsilon}{\mu}}\label{eq63}
\end{align}
in the limit $\mu\rightarrow\infty$, $\varepsilon\rightarrow\infty$, with $M$ being finite. In other words, a PEMC is a very specific limiting case of a bi-isotropic medium with a strong response. As pointed out by Hehl and co-workers \cite{Hehl2008}, Cr${}_2$O${}_3$ is a naturally occurring crystal with a weak nonreciprocal cross-polarisability. They alluded to the close analogy of such an electromagnetic response that of the PEMC as well as the Tellegen medium and the axion field in particle physics. This connection is also discussed in Ref.~\cite{Zhang2015}\\

PEMC materials can be seen as the dual transform of a PEC by a finite duality transformation angle $\theta$. Transforming the PEC-boundary conditions
\begin{align}
\mathbf{n}\cdot{\mathbf{B}}^\star=\mathbf{n}\cdot(-\sin(\theta){\mathbf{D}}+\cos(\theta){\mathbf{B}})=&0\\
\mathbf{n}\times{\mathbf{E}}^\star=\mathbf{n}\times(\sin(\theta){\mathbf{H}}+\cos(\theta){\mathbf{E}})=&0
\end{align}
directly gives Eqs.~(\ref{eq03}) and (\ref{eq04}) if the identification 
\begin{equation}\label{MDef}
M=\cot(\theta)
\end{equation}
is made. This means that the PEC case corresponds to $\theta=0$ and the PMC case to $\theta=\pi/2$, with all other cases appearing for intermediate angles in the range $(0,\pi/2)$ as shown in Fig.~\ref{fig:fig5}  \\
\begin{figure}[h]
\centering
	\begin{tikzpicture}
	\draw (50mm,0) arc (0:90:50mm);
	\draw[->, cyan, very thick] (0,0)  -- (5,0)node[below left, black]{PMC};
	\draw[->, gray, very thick] (0,0)  -- (0,5)node[above right, black]{PEC};
	\draw[->, cyan!50!gray, very thick] (0,0)  -- (2.5,4.33) node[above right, black]{PEMC};
	\draw (0,0) --(0,3)  node[below right=1em,black]{\huge $\theta$} arc(90:60:30mm)  --(0,0);
	\end{tikzpicture}
	\caption{Relation between PEC, PEMC, PMC and the duality angle $\theta$.}
	\label{fig:fig5}
\end{figure}
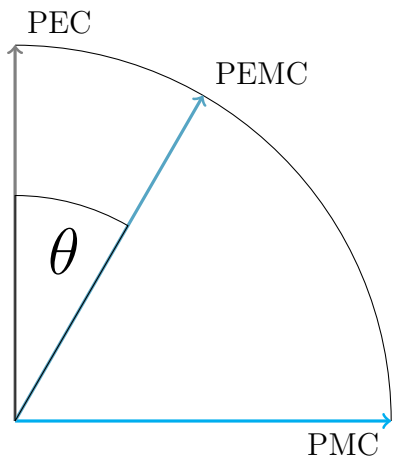\\

An assembly involving PECs can hence easily be transformed to a PEMC setup by means of duality transformation (see Fig.~\ref{fig:fig5}). Note that a global duality transformation is supposed not to change any physical results. This leads immediately to the prediction that the Casimir force between two identical PEMC plates is the same as for two plates of PECs, as will be explicitly verified in Section \ref{CasimirForceSection}.\\

\section{The Green's tensor of two PEMC plates}  

In order to apply our general result \eqref{eq54} to two PEMC plates, we first need to find the respective Green's tensor. 

\subsection{General structure of the Green's tensor} 
The reflection properties of a nonreciprocal plate are described by four reflection coefficients $r_{ss}$, $r_{sp}$, $r_{ps}$, $r_{pp}$ corresponding to all possible combinations of the polarisation directions ($s$ or $p$) of the incoming and outgoing light. Here the index $p$ denotes an electric field polarisation parallel to the normal vector of the surface [transverse-magnetic (TM) polarisation], while $s$ indicates perpendicular polarisation [transverse-electric (TE) polarisation]. The reflected wave $\B{v}_\text{refl}$ corresponding to a general incident wave $\B{v}_\text{inc}$ at a boundary described by these four coefficients can therefore be represented as a matrix multiplication:  
\begin{equation}
\mathbf{v}_{\text{refl}}=R\cdot \mathbf{v}_{\text{inc}}=\begin{pmatrix}
r_{ss}& r_{sp}\\r_{ps}& r_{pp}
\end{pmatrix}
\cdot\begin{pmatrix}
v_s\\
v_p
\end{pmatrix}
\end{equation}
A setup consisting of two plates is considered as a three-layer system,   
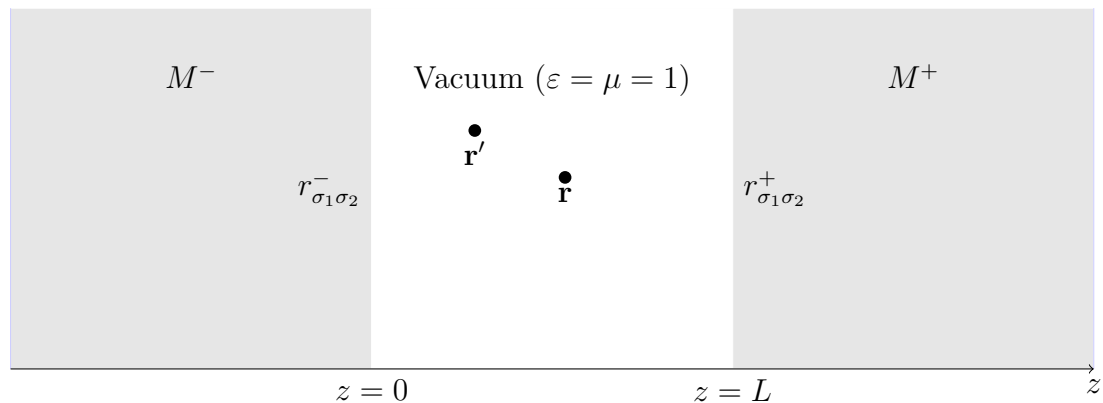
\begin{figure}[h]
\begin{tikzpicture}[scale=1.2]
\filldraw[fill=gray!20, draw=white!50!white] (-4,0) rectangle (4,4);
\draw[blue!20] (-4,0) -- (-4,4);
\filldraw[white!20, draw=white!50!white] (0,0) rectangle (4,4);
\filldraw[fill=gray!20, draw=white!50!white] (4,0) rectangle (8,4);
\draw[blue!20] (8,0) -- (8,4);
\fill (2.143,2.1235) circle[radius=2pt] node[below]{$\mathbf{r}$};
\fill (1.143,2.6435) circle[radius=2pt] node[below]{$\mathbf{r}'$};
\fill[white!20] (2.0,3.5) circle[radius=2pt] node[below,black]{Vacuum ($\varepsilon=\mu=1$)};
\fill[gray!20] (6.0,3.5) circle[radius=2pt] node[below,black]{$M^+$};
\fill[gray!20] (-2.0,3.5) circle[radius=2pt] node[below,black]{$M^-$};
\fill[blue!20] (4.0,2.0) circle[radius=0pt] node[right,black]{$r^+_{\sigma_1\sigma_2}$};
\fill[blue!20] (0,2.0) circle[radius=0pt] node[left,black]{$r^-_{\sigma_1\sigma_2}$};
\fill[blue!20] (0,0) circle[radius=0pt] node[below,black]{$z=0$};
\fill[blue!20] (4.0,0) circle[radius=0pt] node[below,black]{$z=L$};
\draw[->] (-4,0) -- (8,0) node[below,black]{$z$};
\end{tikzpicture}
\caption{Three-layer system with perfectly reflecting, cross-polarising boundary surfaces at $z=0 $ and $z=L$}
\label{fig:fig4}
\end{figure}
where we require the Green's tensor for all positions in the middle layer. This consists of waves travelling from $\mathbf{r}$ to $\mathbf{r}'$ and being reflected any number of times, which can be elegantly taken into account by means of a Neumann series. For matrices $R^\pm$ representing two plates being located at $z=0$ ($R^-$ ) and $z=L$ ($R^+$ ) respectively we define
\begin{align}
D^\pm_{\sigma_i\sigma_j}&=\left[\sum^\infty_\text{n=0}(R^\pm\cdot R^\mp)^n\cdot(e^{-2\mathrm{i}k^\perp L})^n\right]_{\sigma_i\sigma_j}=\left(\mathbb{I}-R^\pm\cdot R^\mp e^{-2\mathrm{i}k^\perp L}\right)_{\sigma_i\sigma_j}^{-1}
\end{align}
with $\sigma_i$, $\sigma_j$ denoting the polarisation directions $s$ and $p$. Using the general form of the Green's tensor we obtain the result
\begin{align}
\label{eq66}\nonumber\tens{G}^{(1)}(\mathbf{r},\mathbf{r}',\omega)=&\frac{1}{8\pi^2}\int\frac{d^2k^{\parallel}}{k^{\perp}}e^{i\mathbf{k}^{\parallel}\cdot(\mathbf{r}-\mathbf{r'})}\\
\times \nonumber&\phantom{=}\, \left[ \sum_{\sigma_1\sigma_2}\mathbf{e}_{\sigma_1+}\cdot R^+\cdot(D^\mp) ^{-1}\cdot R^-\cdot\mathbf{e}_{\sigma_2+}e^{\mathrm{i}k^\perp(2L+z-z')}\right.\\
\nonumber&\phantom{=}\, +\sum_{\sigma_1\sigma_2}\mathbf{e}_{\sigma_1-}\cdot R^-\cdot(D^\pm)^{-1}\cdot R^+\cdot\mathbf{e}_{\sigma_2-}e^{\mathrm{i}k^\perp(2L-z+z')}\\
\nonumber&\phantom{=}\,+\sum_{\sigma_1\sigma_2}\mathbf{e}_{\sigma_1-}\cdot R^-\cdot(D^\mp)^{-1}\cdot\mathbf{e}_{\sigma_2+}e^{\mathrm{i}k^\perp(z+z')}\\
& \phantom{=}\,+\left.\sum_{\sigma_1\sigma_2}\mathbf{e}_{\sigma_1+}\cdot R^+\cdot(D^\pm)^{-1}\cdot\mathbf{e}_{\sigma_2-}e^{\mathrm{i}k^\perp(2L-z-z')}\right].
\end{align}

with

\begin{align}
\mathbf{e}_{s}^\mp&=\mathbf{e}_{s}^\pm=\mathbf{e}_{k^\parallel}\times\mathbf{e}_z, &
\mathbf{e}_{p}^\pm&=i/|k|(k^\parallel\mathbf{e}_z\pm k^\perp\mathbf{e}_{k^\parallel}) &
\mathbf{k}&=k^\parallel\mathbf{e}_{k^\parallel}+k^\perp\mathbf{e}_{k^\perp}.
\end{align}
Note that the matrix multiplication is performed in $(s,p)$-space. The Green's tensor's spatial components are obtained by the outer product of the respective polarisation vectors. In this expression the first two terms account for an even number of multiple reflections between $\mathbf{r}$ and $\mathbf{r'}$, while odd numbers of reflections contribute to the final two terms. Similarly to the case of reciprocal materials \cite{Buhmann2012a}, the terms representing an odd number of reflections do not contribute to the Casimir force because they can be seen as part of the self energy of one plate.

\subsection{PEMC reflection matrices}
The boundary conditions \eqref{eq03} and \eqref{eq04} for the fields lead to polarisation-mixing effects at a PEMC boundary. In terms of the magnetoelectric constants these reflection coefficients are given by \cite{Fuchs2017}  
\begin{align}
r_{ss}&=\frac{(  k_1^\perp-\mu k^\perp_2)\Omega_\varepsilon-k_1^\perp k_2^\perp\xi^2}{(  k_1	^\perp-\mu k^\perp_2)\Omega_\varepsilon+k_1^\perp k_2^\perp\xi^2},\\
r_{ps}&=\frac{-2\mu  k_1^\perp k_2^\perp\xi}{(k_1^\perp-\mu k^\perp_2)\Omega_\varepsilon+k_1^\perp k_2^\perp\xi^2}=r_{sp},\\
r_{pp}&=\frac{(  k_1^\perp-\left(\varepsilon-\frac{\xi^2}{\mu}\right)k^\perp_2)\Omega_\mu-k_1^\perp k_2^\perp\xi^2}{( k_1^\perp-\left(\varepsilon-\frac{\xi^2}{\mu}\right)k^\perp_2)\Omega_\mu+k_1^\perp k_2^\perp\xi^2},
\end{align}
with $\Omega_\mu= \mu(  k_1^\perp+\mu k^\perp_2)$ and $\Omega_\varepsilon= \mu[  k_1^\perp+\left(\varepsilon-{\xi^2}/{\mu}\right)k^\perp_2]$ and $k^\perp_i$ representing the perpendicularly polarised part of the wave on the two sides of the plate. In the PEMC-limit, with all response functions going to infinity and $M=\sqrt{\varepsilon/\mu}$, one obtains in matrix form:
\begin{align}
R=
\begin{pmatrix} 
r_{ss}& r_{sp}\\r_{ps}& r_{pp}
\end{pmatrix}=\frac{1}{1+M^2}
\begin{pmatrix} 
{1-M^2}& {-2M}\\{-2M}& {M^2-1}
\end{pmatrix}
\end{align}
which is independent of the incoming wave vector. 

Introducing the corresponding duality transformation angle $\theta$ via Eq.~\eqref{MDef}, one obtains for two plates;
\begin{align}
R^\pm=
\begin{pmatrix} 
-\cos(\theta^\pm)&\sin(\theta^\pm)\\ \sin(\theta^\pm)& \cos(\theta^\pm)
\end{pmatrix}.
\end{align}
with $\theta^\pm$ being the respective duality transformation angle that defines the properties of each plate. We can now also calculate the corresponding multiple-reflection contributions to obtain
\begin{align}
(D^\pm)^{-1}=&\frac{b}{1-2b\cos(2\delta)+b^2}\begin{pmatrix}
b-\cos(2\delta)&\sin(2\delta)\\
-\sin(2\delta)& b-\cos(2\delta)
\end{pmatrix}\\
(D^\mp)^{-1}=&\frac{b}{1-2b\cos(2\delta)+b^2}\begin{pmatrix}
b-\cos(2\delta)&-\sin(2\delta)\\
\sin(2\delta)& b-\cos(2\delta)
\end{pmatrix}
\end{align}   
with $b=e^{-2\mathrm{i}k^\perp L}$ and $\delta=\theta^+-\theta^-$.
\section{Casimir force between two PEMC plates}\label{CasimirForceSection}

In order to solve the Green's tensor integral we introduce polar coordinates $(k^\parallel,\varphi)$ for the two-dimensional integral over $\mathbf{k}^\parallel$. This simplifies the calculation considerably because the reflection matrices as well as $D^\pm$ and $D^\mp$ do not depend on $\varphi$, the angular dependence appears only in the dyadic product of the polarisation vectors. These can be straightforwardly be integrated as in
\begin{gather}
\label{eq70}\nonumber\int_0^{2\pi} d\varphi\ \mathbf{e}^\pm_s\otimes\mathbf{e}^\mp_s=\int_0^{2\pi} d\varphi\ \mathbf{e}^\pm_s\otimes\mathbf{e}^\pm_s\\
=\int_0^{2\pi} d\varphi\ (\mathbf{e}_{k^\parallel}\times\mathbf{e}_z)\otimes(\mathbf{e}_{k^\parallel}\times\mathbf{e}_z)
=\pi(\mathbf{e}_x\otimes\mathbf{e}_x+\mathbf{e}_y\otimes\mathbf{e}_y)\\
\int_0^{2\pi} d\varphi\ \mathbf{e}^\pm_p\otimes\mathbf{e}^\pm_p=-\frac{\pi c^2}{\xi^2}\left[2k^{\parallel2}\mathbf{e}_z\otimes\mathbf{e}_z+k^{\perp 2}(\mathbf{e}_x\otimes\mathbf{e}_x+\mathbf{e}_y\otimes\mathbf{e}_y)\right]
\end{gather}
and so on. \\
We can compute a force $d\mathbf{F}/dA$ per unit area from the stress tensor via Eq.~\eqref{eq8}. Making use of the fact that $d\mathbf{A}\parallel\mathbf{e}_z$, we have; 
\begin{align}
\label{eq71}\nonumber \mathbf{f}=&\frac{d\mathbf{F}}{dA}=-\frac{1}{A}\int_{\partial V}d\mathbf{A}\cdot\mathbf{T}\\
=&-\frac{\hbar}{2\pi}\int_0^\infty d\xi\sum_{j=x,y,z}^{}(\mathbf{T}_{jz}\mathbf{e}_j)\Big|_{z=L}
\end{align}
where $\left<\mathbf{T}\right>$ is given by (\ref{eq54}). The symmetry of the problem requires that $\mathbf{f}$ has no $x$- or $y$- components which can indeed be seen from the fact that no combination of the polarisation vectors yields a $x$-$z$- or $y$-$z$-component when integrated over $\varphi$. We will hence suppress the fact that $\mathbf{f}$ is a vector and just calculate its absolute value. \\
 We now insert our obtained Green's tensor (\ref{eq66}) into (\ref{eq54}) and observe that the contributions from the terms containing curls equal the contributions from those without. After setting $\kappa=\mathrm{i}k^\perp$, transforming to polar co-ordinates $k^\parallel=\kappa\cos(\phi)$, $\xi/c=\kappa\sin(\phi)$ and carrying out the trivial angular integration we get
\begin{equation}
 f=\frac{\hbar c}{4\pi^2}\int_0^\infty \!\!\!d\kappa\ \kappa^3\  e^{-2\kappa L}\sum_{\sigma_1\sigma_2}^{}\Big[\mathbf{e}_{\sigma_1}\cdot( R^+\cdot (D^\mp)^{-1}\cdot R^-+R^-\cdot (D^\pm)^{-1}\cdot R^+)\cdot\mathbf{e}_{\sigma_2}\Big]
 \end{equation}
This generalisation of Lifshitz's formula for planar systems \cite{E.M.Lifshitz1956} agrees with results for reciprocal polarisation-mixing plates such as gratings \cite{Lambrecht2008, Contreras-Reyes2010, Bender2014}. Remarkably, the result is hence insensitive to the fact that the plates are non-reciprocal at this level. 

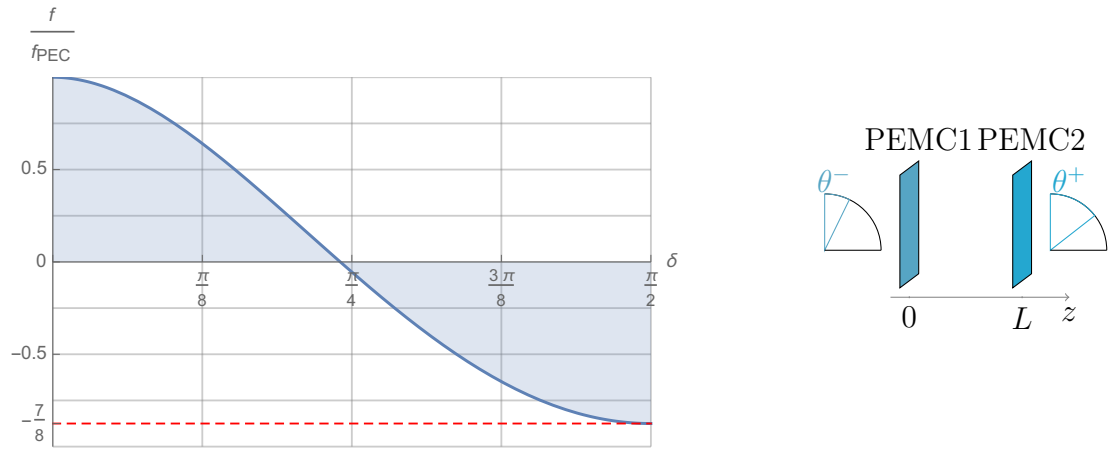
\begin{figure}[h]
\label{fig1}
\centering
\begin{minipage}{0.55\textwidth}
\includegraphics[scale=0.7]{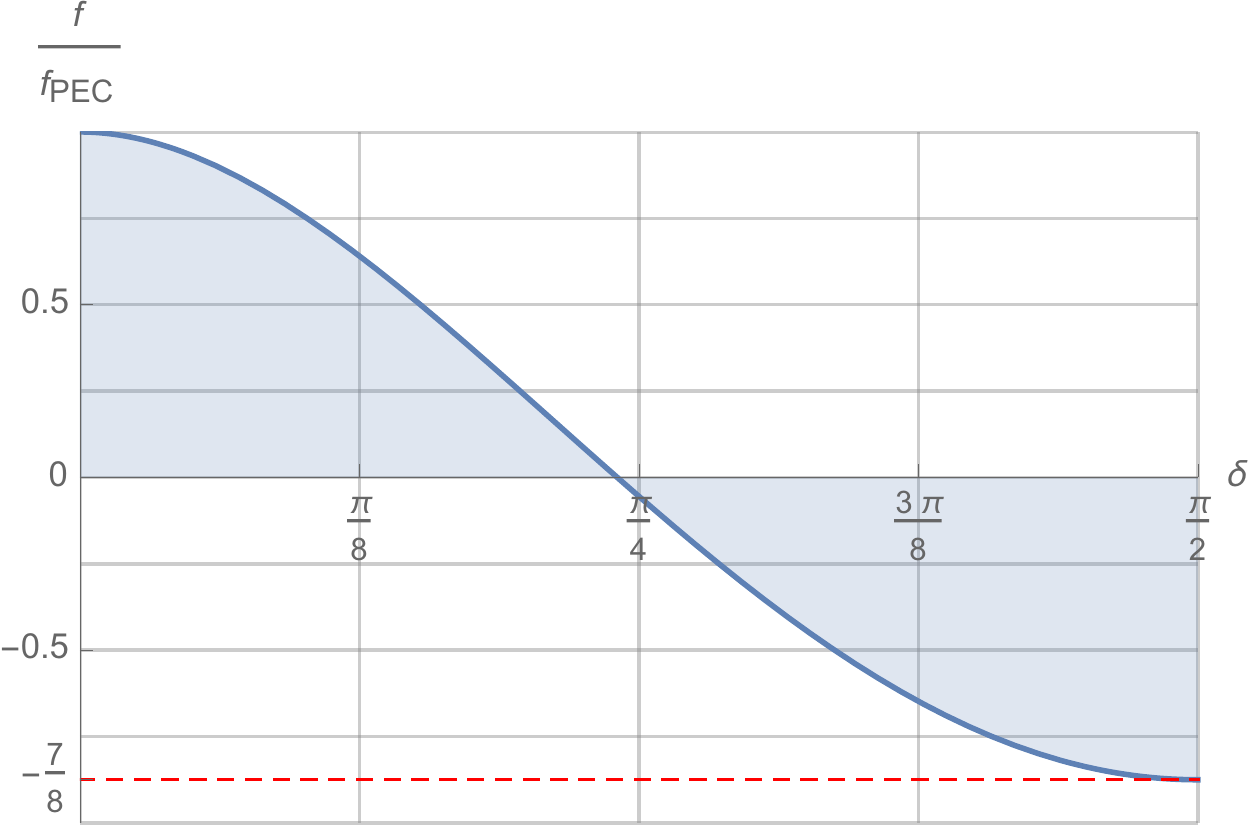}
\end{minipage}
\hfill
\begin{minipage}{0.325\textwidth}
\begin{tikzpicture}[xscale=0.25,yscale=0.25]
	\node[cyan!50!gray] at (-3.5,4.75){$\theta^-$};
	\draw (-1,1) arc (0:90:30mm);
	\draw[cyan!50!gray] (-4,4) arc(90:64:30mm)  -- (-4,1) -- (-4,4);
	\draw (-4,1) -- (-1,1);
	\filldraw[fill=cyan!50!gray, draw=black!] (1,-0.25)--(0,-1)--(0,5)--(1,5.75) node[above, black]{PEMC1}--(1,-0.25) ;
	\filldraw[fill=cyan!65!gray, draw=black!]  (7,-0.25)--(6,-1)--(6,5)--(7,5.75)node[above, black]{PEMC2}--(7,-0.25);
	\draw[->,gray, very thin] (-0.5,-1.5) -- (9,-1.5) node[below, black]{$z$};
	\draw[gray, very thin] (0.5,-1.55) -- (0.5,-1.45) node[below, black]{$0$};
	\draw[gray, very thin] (6.5,-1.55) -- (6.5,-1.45) node[below, black]{$L$};
	\draw (8,4) arc (90:0:30mm);
	\draw[cyan!65!gray] (8,4) arc(90:38:30mm)  -- (8,1) -- (8,4);
	\draw (8,1) -- (11,1);
	\node[cyan!65!gray] at (9,4.75){$\theta^+$};
	\end{tikzpicture}
\end{minipage}
\caption{Casimir force between two PEMC plates normalised to the original result of Casimir in terms of their duality phase shift $\delta=\theta^+-\theta^-$}
\end{figure}

Substituting $x=\kappa L$ and performing the matrix multiplications we find the following integral
\begin{align}
\nonumber f=&-\frac{\hbar c}{\pi^2L^4}\int_0^\infty dx\ x^3 \frac{e^{2x}\cos(2\delta)-1}{1-2e^{2x}\cos(2\delta)+e^{4x}}\ \ \\
 =&-\frac{\hbar c}{\pi^2L^4}\int_0^\infty dx\ x^3\left(\frac{\frac 12 e^{2x}(e^{2i\delta}+e^{-2i\delta})}{(1-e^{2x}e^{2i\delta})(1-e^{2x}e^{-2i\delta})}-\frac{1}{(1-e^{2x}e^{2i\delta})(1-e^{2x}e^{-2i\delta})}\right)
\end{align}
which can be analytically integrated to finally obtain our main result
\begin{align}\label{MainResult}
\mathbf{f}(\theta^+,\theta^-)=-\frac{3\hbar c}{8\pi^2L^4}\text{Re}\left(\text{Li}_4\left[e^{2i(\theta^+-\theta^-)}\right]\right)\mathbf{e}_z
\end{align}
where we have made use of the polylogarithm function $\text{Li}_n(z)=\sum_{k=1}^{\infty}z^k/k^n$. Our result \eqref{MainResult} immediately demonstrates the duality invariance of the Casimir force; it only depends on the difference of the PEMC angles, so a globally applied duality transformation does not change the Casimir force. Thus we may write $\mathbf{f}(\theta^+,\theta^-)=\mathbf{f}(\theta^+-\theta^-)=\mathbf{f}(\delta)$. We can easily check that this is indeed compatible with the results of Casimir and Boyer via
\begin{equation}
\text{Re}\, \text{Li}_4(e^{i\phi})=\sum_{k=1}^\infty\frac{\cos^k (k\phi)}{k^4}=\frac{\pi^4}{90}-\frac{\pi^2 \phi^2}{12}+\frac{\pi\phi^3}{12}-\frac{\phi^4}{48}
\end{equation}
where we used de Moivre's identity followed by formula 27.8.6(3) of \cite{Abramowitz1970}, giving
\begin{equation}\label{MainResultSimp}
\mathbf{f}(\delta)=-\frac{\hbar c}{8\pi^2L^4}\left[ \frac{\pi^4}{ 30}-\delta^2 (\pi-\delta)^2 \right]\mathbf{e}_z
\end{equation}
Then we obtain the special cases of Casimir ($\delta=0$)
\begin{align}
\mathbf{f}(0)=-\frac{\hbar c}{240\pi^2L^4}\mathbf{e}_z
\end{align}
and Boyer ($\delta=\pi/2$)
 \begin{equation}
 \mathbf{f}(\pi/2)=\frac{7}{8}\cdot \frac{\hbar c}{240\pi^2L^4}\mathbf{e}_z
\end{equation}
We show the results for intermediate angles in Fig.~\ref{fig1}.
It is seen that there is some value $\delta_\text{crit}$ for which there is no Casimir force, given by solving 
\begin{equation}
\frac{\pi^4}{ 30}-\delta_\text{crit}^2 (\pi-\delta_\text{crit})^2=0  \end{equation}
giving;
\begin{equation}
\delta_\text{crit}=\frac{\pi}{2} \Bigg( 1-\sqrt{1-2 \sqrt{\frac{2}{15}}}  \Bigg) \approx 0.96 \cdot \frac{\pi}{4}
\end{equation}
  It is also interesting to notice that the following
  \begin{equation}
\int_0^{\pi/2}d\delta \mathbf{f}(\delta)=0,
\end{equation}
so even though the force is not symmetric around the central angle $\delta=\pi/4$, the enclosed areas to the left and the right of the zero-force angle $\delta_\text{crit}$ are equal. Thus our result represents a sum rule for the Casimir force for PEMCs; the sum of Casimir forces over the entire PEMC parameter space is zero.

\section{Conclusion}
In order to calculate the Casimir force between two PEMC plates we have constructed the Green's tensor for two nonreciprocal plates in terms of their reflection properties. The result is duality invariant as well as it is compatible with the theorem derived by Kenneth and Klich that the Casimir force between identical bodies is always attractive \cite{Kenneth2002}. It also extends to nonreciprocal media the prediction that the Casimir force between two plates of any possible material will fall in between the results of Casimir and Boyer \cite{Henkel2017}, which had thus far only been shown for magnetodielectrics. The desired Green's tensor is hence also applicable for different lossless material classes. In particular of the focus might go to chiral perfectly reflecting chiral materials ($\xi=-\zeta$ in terms of material constants) to explore the full parameter space of the Casimir effect. \\
For more realistic scenarios of course the corrections due to imperfect reflection are of high interest. For these cases the derived PEMC case can be viewed as a theoretical upper limit for the Casimir force since we assumed the reflection coefficients as well as the PEMC parameter to be frequency independent. In less idealised cases one would expect the resulting Casimir force to lie somewhere 'under the curve' for the respective value of $\theta$.  

\section*{Acknowledgements}
We would like to thank F. Hehl for inspiring this work and S. Fuchs, F. Lindel and A. Sihvola for stimulating discussions. The authors acknowledge financial support from the Deutsche Forschungsgemeinschaft via grants BU1803/3-1 and GRK 2079/1. R.B. and S.Y.B additionally acknowledge support from the Alexander von Humboldt foundation, and S.Y.B. acknowledges support from the Freiburg Institute of Advanced Studies (FRIAS).

\end{document}